\documentstyle[12pt,rotate]{article}
\begin{document}
\vspace{0.4in}
\begin{center}
{\large\bf Phonon-Like Excitations of the Instanton Liquid
}\\[0.5cm]
{S.V. Molodtsov\footnote{molodtsov@vitep5.itep.ru}, 
A.M. Snigirev\footnote{snigirev@lav.npi.msu.su}, 
G.M. Zinovjev}\footnote{gezin@hrz.uni-bielefeld.de, 
gezin@ap3.bitp.kiev.ua}$^{,4}$\\[0.5cm]
\end{center}
{\small\it $^1$ 
State Research Center, Institute of Theoretical and Experimental Physics,
Moscow, Russia\\
$^2$ Nuclear Physics Institute, Moscow State University,
Moscow, Russia\\
$^3$Fakult{\"a}t f{\"u}r Physik, Universit{\"a}t Bielefeld,
Bielefeld, Germany\\
$^4$Bogolyubov Institute for Theoretical Physics,
National Academy of Sciences of Ukraine.
}

\thispagestyle{empty}

\begin{abstract}
The phonon-like excitations of (anti)-instanton liquid ($\bar II$) due to
adiabatic variations of vacuum wave functions are studied in this
paper. The kinetic energy term is found and proper effective 
Lagrangian for such excitations is evaluated.The properties of their 
spectrum, while the corresponding masses are defined by $\Lambda_{QCD}$
with prevailing chromoelectric component, are investigated based on the
phenomenology of QCD vacuum already developed. 
\end{abstract}
\vspace{0.5cm}
PACS: 11.15 Kc, 12.38-t, 12.38-Aw
\\
\\
Today's great splash of interest in instanton-based models was provoked by
recent attempts to advance our understanding of QCD behaviour at very high
densities \cite{0}. It was argued there that instantons are able to
generate a gap quite suitable to revive the mechanism of the 'colour
superconductor' in QCD for quantitative utility. In fact, the instantons and,
in particular, the model of (anti)-instanton liquid correctly seize many 
nonperturbative phenomena and vacuum features as chiral symmetry breaking, 
gluon condensate, topological susceptibility \cite{1},\cite{2}. However, 
pushing forward the detailed comprehension of QCD principal issues, new
questions appear that can be addressed to the instanton liquid model for 
its own sake. We are focused on one of those in this paper.

It is usually supposed the corresponding functional integral in this
approach is saturated by quasi-classical configurations close to the exact 
solutions of the Yang-Mills equations (the Euclidean solutions called the
(anti)-instantons) and the wave function of vacuum being homogeneous in
metric space is properly reproduced by averaging over their collective
coordinates. In the $\bar II$ liquid approach one takes the superposition 
ansatz of the pseudo-particle (PP) fields as a relevant approximation to
the 'genuine' vacuum configuration 
\begin{equation}
\label{1} 
A_\mu(x)=\sum_{i=1}^N A_\mu(x;\gamma_i)~.
\end{equation}
Here $A_\mu(x;\gamma_i)$ denotes the field of a singled (anti)-instanton
in singular gauge with $4N_c$ (for $SU(N_c)$ group) coordinates  
$\gamma=(\rho,~z,~\Omega)$, of the size $\rho$ with the
coordinate of its centre $z$ and $\Omega$ as its colour orientation. 
Then,
\begin{equation}
\label{2}
A^{a}_\mu(x;\gamma)=\frac {2}{g}~\Omega^{ab} \bar \eta_{b\mu\nu}
\frac{y_\nu}{y^2}\frac{\rho^2}{y^2+\rho^2}~,~~y=x-z~,
\end{equation}
where $\eta$ is the t'Hooft symbol \cite{3}.
For anti-instanton $\bar\eta\to\eta$ (making the choice of the singular
gauge allows us to sum up the solution preserving the asymptotic
behaviour). For simplicity we shall not introduce the different symbols
for instanton and anti-instanton, and then in the superposition of
Eq.(\ref{1}) $N$ becomes the total number of PP in the system 4-volume $V$ 
with the density $n=N/V$.

The 'liquid' is introduced by the distribution function over instanton sizes 
and in one-loop approximation it is taken as the following
\begin{equation}   
\label{3}
\mu(\rho)= \rho^{-5} \widetilde\beta^{2 N_c}
e^{-\beta(\rho)-\nu \rho^2/\overline{\rho^2}}~.
\end{equation}   
As known, the respective parameters are  defined by the QCD
renormalization scale $\Lambda_{QCD}$ together with the constant $\xi$ 
characterizing effective infrared repulsion of the PP's in the model of 
Ref.\cite{2}, for example, 
\begin{equation}   
\label{nom}
\left(\overline{\rho^2}\right)^2=\frac{\nu}{\beta \xi^2 n}~,~~
\xi^2=\frac{27}{4}\frac{N_c}{N_c^{2}-1} \pi^2~,~~
\nu=\frac{1}{2}(b-4)~,~~b=\frac{11}{3} N_c~,
\end{equation}
with the Gell-Mann--Low beta function 
\begin{equation}
\label{beta}
\beta(\rho)=-\ln C_{N_c}-b \ln(\Lambda \rho)~,~~
\Lambda=\Lambda_{\overline{MS}}=0.92 \Lambda_{P.V.}~,
\end{equation} 
and constant $C_{N_c}$ depending on the regularization scheme, in particular,
$$ C_{N_c}\approx\frac{4.6~\exp(-1.37 N_c)}{\pi^2(N_c-1)!(N_c-2)!}~.
$$
The parameters $\beta$ and $\widetilde \beta$ are fixed at the characteristic
scale $\bar\rho$, i.e.
$$
\beta=\beta(\bar\rho)~,~~
\widetilde \beta=\beta -\ln C_{N_c}~.
$$
The distribution $\mu(\rho)$ has obvious physical meaning, namely,
the quantity $d^4x~d \rho~\mu(\rho)$ is proportional to the probability to 
find an instanton of size $\rho$ in some point of a volume element $d^4 x$. 
At small $\rho$ the behaviour of distribution function is stipulated by the 
quantum-mechanical uncertainty principle preventing a solution being 
compressed at a point (radiative correction). At large $\rho$ the constraint 
comes from repulsive interaction between the PP's which is amplified with
(anti)-instanton size increasing.

In what follows, we study the excitations of $\bar II$ liquid induced
by the adiabatic dilatational deformations of instanton solutions. We
show the corresponding excitations display a particle-like ('phonon')
behaviour, developing the mass spectrum settled by $\Lambda_{QCD}$ and
with the chromoelectric component prevailing. It looks quite plausible that
incorporating light quark condensate would allow us to describe the light
hadrons on the correct mass scale, and to provide more insight into the
confinement mechanism.

The guiding idea of selecting deformation originates from transparent
observation. The deformations measured in the units of action 
$\frac{dq~dp}{2\pi\hbar}$ (here $q,~p$ are the generalized coordinate and
momentum) have only physical meaning. However, the instantons are 
characterized by 'static' coordinates $\gamma$ and, therefore, need to
appoint the momenta conjugated. It looks quite natural for the variable $\rho$ 
to introduce $\dot\rho=d \rho/dx_4$ as a conjugated momentum. Then we should 
include additional degrees of freedom induced by the variation of instanton 
size, and under sufficiently slow variation of its size the additional terms 
occur in strength gluon tensor
\begin{equation}
\label{4}
G'^a_{\mu\nu}=G^a_{\mu\nu}+g^a_{\mu\nu}~.
\end{equation}
The first term corresponds to the contribution generated by the 
instanton profile 
$$ G^a_{\mu\nu}=-\frac{8}{g}\frac{\rho^2}{(y^2+\rho^2)^2} 
\left (\frac12 \bar\eta_{a\mu\nu} + \bar\eta_{a\nu\rho}
\frac{y_\mu y_\rho}{y^2}
-\bar\eta_{a\mu\rho}\frac{y_\nu y_\rho}{y^2}
\right)~,
$$
and in adiabatic approximation the corrections have the form
$$ g^a_{4i}\approx\frac{\partial A^a_i}{\partial\rho}\dot\rho=
\frac4g\bar\eta_{ai\nu}~\frac{y_\nu~\rho}{(y^2+\rho^2)^2}~\dot\rho,~~
g^a_{ij}=0~,~~g^a_{i4}=-g^a_{4i}~,~~i, j=1,2,3~.
$$
Here we restrict ourselves with the superposition ansatz (\ref{1}) and 
neglect the terms of order $O(\dot\rho^2).$
The adiabatic constraint $g^a_{\mu\nu}\ll G^a_{\mu\nu}$
 means the variation of instanton size is much
smaller than characteristic transformation scale of the PP field, 
$\dot\rho\ll O(1).$ Then calculating the corrections for the 
action, it is reasonable to take out $\dot\rho$ beyond the integral and
one instanton contribution to the action turns out
\begin{equation}
\label{5}
S=\frac14 \int d^4 x~ G'^2_{\mu\nu}\simeq\frac{8\pi^2}{g^2}+C~\dot\rho+
\frac{\kappa_{s.t.}(\rho)}{2}~\dot\rho^2 ,
\end{equation}
where $\dot\rho$ should be taken as the mean rate of slow solution
deformation at characteristic instanton life-time $\sim \rho$. For
simplicity, one may take it in the centre of instanton $\dot\rho (z).$ 
The constant $C$ coming from the interference term should be equal to zero
as a result of space homogeneity and for the 'kinematical' 
$\kappa$-term we have \footnote{In a regular gauge the result is the same.}
\begin{equation}
\label{6}
\frac{\kappa_{s.t.}(\rho)}{2}=
\frac12 \int d^4 x~ g^2_{4i}=
\frac{3\pi^2}{g^2}~.
\end{equation} 
The overt $\rho$ dependence of $\kappa$ is lacking because of the scale 
invariance. It arises with the renormalization of the coupling constant. 
Dealing with definition (\ref{1}) we have considered only 
the corrections induced by the variation of strength tensors in line with
anzatz (\ref{4}), but not those
resulting from a possible variation of fields (\ref{2}). Bearing in mind
the form of potentials in regular ($r.g.)$ and singular $(s.g.)$ gauges
\begin{eqnarray}
\label{a1}
A^{a}_\mu&=&\frac{1}{g} \eta_{a\mu\nu}~
\partial_\nu \ln(y^2+\rho^2),~~~~~~{\mbox{ r.g.}}
\nonumber\\ [-.2cm]
\\ [-.25cm]
A^{a}_\mu&=&-\frac{1}{g} \bar \eta_{a\mu\nu}~
\partial_\nu \ln\left(1+\frac{\rho^2}{y^2}\right),~~~~~~{\mbox{ s.g.}}
\nonumber
\end{eqnarray}
we find the adiabatic corrections $A_\mu^{'}=A_\mu + a_\mu$ as follows: 
\begin{equation}
\label{a2}
a^{a}_\mu=\frac{2}{g} \eta_{a\mu 4}~
\frac{\rho}{y^2+\rho^2}~\dot\rho,~~~~~~{\mbox{r.g.}}
\end{equation}
as for the profiles (\ref{a1}), the substitution $\eta \to-\bar\eta$
brings about the transition from regular gauge to a singular one. 
Consequently, it becomes possible to calculate the correction for the
chromoelectric field (in the singular gauge, for clarity)
\begin{equation}
\label{a5}
\epsilon^{a}_i(p.)=g^a_{4 i}=\frac{4}{g}~
\frac{\rho~ y_4~\dot\rho}{(y^2+\rho^2)^2}~\delta_{a i}~,
\end{equation}
(p. means the potential part here), and the correction for the
chromomagnetic field is
\begin{equation}
\label{a6}
h^{a}_i=\frac{1}{2}\varepsilon_{i j k}~ g^a_{j k}=-\frac{4}{g}~
\frac{\rho~\dot\rho}{(y^2+\rho^2)^2}~\varepsilon_{a i j}~ y_j~.
\end{equation} 
Chromomagnetic field contributes to the action as
\begin{equation}
\label{a7}
\frac12 \int d^4 x~ h^a_{i}h^a_{i}=
\frac{4\pi^2}{g^2}~\dot\rho^2,
\end{equation} 
and the contribution of chromoelectric field 
($~~\epsilon^{a}_i=\epsilon^{a}_i(p.)+\epsilon^{a}_i(s.t.)~~$) 
is given in the following form
\begin{equation}
\label{a8}
\frac12 \int d^4 x~ \epsilon^a_{i}\epsilon^a_{i}=
\frac{12\pi^2}{g^2}~\dot\rho^2~.
\end{equation} 
Finally, summing up both contributions to the kinematical $\kappa$-term 
results in
\begin{equation}
\label{a9}
\kappa=\frac{32\pi^2}{g^2}~.
\end{equation} 
The comparison of Eq.(\ref{a7}) and Eq.(\ref{a8}) allows us to conclude that 
the chromoelectric fields are influenced mainly by dilatational deformations. 

The formulae obtained characterize the deformed instanton, and one can see
the adiabatic variation of the PP solution does not change the action
considerably. A small contribution of order $O(\dot\rho^2)$ only is added.
Turning to the $\bar I I$ liquid, let us rewrite the PP ensemble formulae in 
the compact form of a density functional (as in statistical mechanics). 
Assuming the number of PP in an ensemble is still 
appropriate to consider them separately, one may rewrite the superposition
Eq.(\ref{1}) in the form of double summing as
\begin{equation}
\label{b1}
A_\mu(x)=\sum^{K}_{i=1}\sum^{\triangle N(\rho_i)}_{j=1}
A_\mu(x;i,\gamma_j)~,
\end{equation}
Here $\triangle N(\rho_i)$ is the PP number of the size $\rho$ in
the interval $(\rho_i,~\rho_i+\triangle\rho)$, $K$ is the number of
partitions within $(\rho_{in},~\rho_{fn})$ and $A_\mu(x;i,\gamma_j)$ is
the (anti)-instanton solution for PP with the calibrated size of 
$\rho \in (\rho_i,~\rho_i+\triangle \rho)$ where $\gamma=(z,~\Omega)$ stands 
for the coordinate of centre and colour orientation of solution.
By definition $\sum^{K}_{i=1}\triangle N(\rho_i)=N.$
The distribution function is taken as
\begin{equation}
\label{b2}
n(\rho)=\frac{\triangle N(\rho)}{\triangle \rho}\frac{1}{V},
\end{equation}
with normalizing $\sum^{K}_{i=1}n(\rho_i)~\triangle\rho~V=N$.
Assuming the limit $\triangle \rho \to 0$ existing one has
$V\int d\rho~n(\rho)=N$. As is known, such an ensemble is described by the
action averaged over the PP coordinates in metric and colour spaces. 
{\footnote {We omit the integration over the instanton sizes because we 
have introduced the collective coordinate of density. In principle, 
the generating functional of the theory should be rewritten in terms 
of functional integration over $D[n(\rho)]$, however, we use here the
distribution function (\ref{3}) found formerly  within
one-instanton approximation.}} Then it is easy to obtain (for noninteracting
system) 
\begin{equation}
\label{n1}
\langle S^{(0)} \rangle=\prod^N_{i=1}\int \frac{d^4 z_i}{V}d 
\Omega_i~S~= \int d^4 x \int d\rho~ n(\rho)
\left\{\frac{8\pi^2}{g^2}+\frac{\kappa(\rho)}{2}~\dot\rho^2\right\}~,
\end{equation}
where the integration over $d^4z$ is carried out in the volume $V$ occupied
by this system,  $d\Omega$ is the measure in colour space
with unit normalization (we omit the argument $\rho$ in the rate
$\dot\rho$ hereafter, hoping it is not misleading). This form of 
$\langle S \rangle$ reflects the homogeneous property of vacuum wave
function assumed, together with taking into account possible adiabatic
variation of the solution profile in various space points. According to
Eq.(\ref{n1}) the deformations at a constant rate additionally favour 
non-interacting PP's. 

The situation becomes crucially different if the interaction is switched on. 
The pair interaction (and the loop corrections) is expected to be of main
importance for our purposes. It is convenient to set the intensity of gluon 
field $G_{\mu\nu}$ generated by any pair of PP's as a sum of field
intensity generated by each separate PP and a term stipulated by
nonlinearity of strength tensors \cite{2}
\begin{eqnarray}
\label{8}
G_{\mu\nu}(a+b)&=&G_{\mu\nu}(a)+G_{\mu\nu}(b)+\triangle G_{\mu\nu}(a,b)~,
\nonumber\\ [-.2cm]
\\ [-.25cm]
\triangle G_{\mu\nu}(a,b)&=&-i\{[a_\mu, b_\nu]+[b_\mu, a_\nu]\}~,\nonumber
\end{eqnarray} 
(in short $G(1+2)=G_1+G_2+G_{12},~G_{\mu\nu}=G^a_{\mu\nu } T^a$ and $T^a$
are the group generators). The cumulative field of the ensemble is given by 
summing up the integral (\ref{n1}) and integrate over all interacting pairs 
of PP's
$$\langle S_{int}\rangle=\int d^4x~\int \frac{d^4 z_1}{V}
d \Omega_1\int \frac{d^4 z_2}{V}d \Omega_2~\frac12~\sum_{12}
\frac14\{G_{12}^2+2 G_1 G_2+2 G_1 G_{12}+2 G_2 G_{12}\}~.
$$
Colour averaging preserves the first term only and with the explicit form
of solution (\ref{2}) we obtain for the partial contribution of every pair
\cite{2}
\begin{equation}
\label{n2}
\frac14\langle G_{12}^2\rangle=\frac{8\pi^2}{g^2}\frac{N_c}{N_c^{2}-1}
\int d^4x\int \frac{d^4 z_1}{V}\int \frac{d^4 z_2}{V}~
\frac{[7 y_1^{2} y_2^{2}-( y_1  y_2)^2]~ \rho_1^{4}\rho_2^{4}}
{ y_1^{4}( y_1^{2}+\rho_1^{2})^2~ y_2^{4}( y_2^{2}+\rho_2^{2})^2}~,
\end{equation}
where $y_i=x-z_i~,~~\rho_i=\rho_i (z_i)~,~~i=1,2.$ If the PP 
sizes do not vary we have for the partial contribution 
$$\langle S_{int}\rangle=
\frac{8\pi^2}{g^2}~\frac{\xi^2}{V}~\rho_1^{2}\rho_2^{2}~.
$$
In the adiabatic approximation the integral (\ref{n2}) is estimated to be
(the contact interaction)
$$\frac14\langle G_{12}^2 \rangle
\simeq
\frac{8\pi^2}{g^2} \frac{\xi^2}{V}
\int \frac{d^4 z}{V}~ 
\rho_1^{2}(z)~\rho_2^{2}(z)~.
$$
Indeed, the adiabatic condition makes it possible to rescale 
the integration 
variable $\frac{dz}{\rho}=d\left (\frac{z}{\rho}\right)+\frac
{d\rho}{dz}\frac{z}{\rho}\approx d\left (\frac{z}{\rho}\right).$
Besides, it is easy to see that the integral is dominated by the 
local vicinity 
of diagonal $\frac{z_1}{\rho_1}\approx \frac{z_2}{\rho_2}$ and the 
distribution function $n(\rho)$ reduces the interaction term $S_{int}$ 
to the following form
\begin{equation}
\label{n3}
\langle S_{int}\rangle=\frac12~\frac{8\pi^2}{g^2}~\xi^2
\int d^4 z \int d\rho_1\int d\rho_2~\rho_1^{2}~n(\rho_1)
~~\rho_2^{2}~n(\rho_2)~,
\end{equation}
remember now $\rho_i~(i=1,2)$ are the functions of $z$.

As known, the loop corrections with pre-exponential factor modify the 
contribution, (\ref{n1}) which is estimated  in our approach as (in terms of
$\beta=\frac{8\pi^2}{g^2}$ as defined by Eq. (\ref{beta})) 
\begin{equation}
\label{n4}
\langle S^{(0)}\rangle=
\int d^4z \int d\rho~ n(\rho)
\left(\frac{\kappa(\rho)}{2}~\dot\rho^2+\beta(\rho)+5 \ln(\Lambda\rho)
-\ln \widetilde \beta^{2N_c}\right)~.
\end{equation}
Finally, one has for the averaged action 
\begin{equation}
\label{n5}
\langle S\rangle=
\langle S^{(0)}\rangle+
\frac12 \beta \xi^2
\int d^4z \int d\rho~ n(\rho)~\rho^2~\int d\rho_2~n(\rho_2)~\rho_2^{2}~.
\end{equation}
With the distribution function (\ref{3}) for stationary $\rho$ we easily 
reproduce the result (\ref{nom}) taking a conventional quantity of averaged 
squared instanton size
$\overline{\rho^2}=\int d\rho~\rho^2~n (\rho)/\int d\rho~n(\rho)~.$ 
Thus, the averaged action is modified to give an effective 
potential in the following form
\begin{eqnarray}
\label{n6}
&&\langle S\rangle=
\int d^4z \int d\rho~ n(\rho)~
\left(\frac{\kappa(\rho)}{2}~\dot\rho^2+U_{eff}(\rho)\right)~,
\nonumber\\ [-.2cm]
\\ [-.25cm]
&&U_{eff}(\rho)=\beta(\rho)+5\ln(\Lambda \rho)-\ln \widetilde \beta^{2N_c}
+\nu~\frac{\rho^2}{\overline{\rho^2}}~.\nonumber
\end{eqnarray}
This result is also used as a suitable approximation herein for the PP's
with deformations if it includes an additional 
integration over volume in the definition of $\bar\rho$ averaged, i.e. 
\begin{equation}
\label{n7}
\overline {\rho^2}=\frac{\int d^4 z\int d\rho~\rho^2~n(\rho)}
{\int d^4 z\int d\rho~n(\rho)}~.
\end{equation}

Everything considered so far concerns the Euclidean action only. In 
order to get a guess for the Minkowski space we notice that in every
temporary slice ($x_4=const$ ) we have a dynamical system (with a natural
transition from $\dot\rho^2$ in Euclidean space to $-\dot\rho^2$ in 
Minkowski one) where, as easily seen, appreciable benefit in action 
could occur. In fact, it takes place if the rates of solution
deformations are self-consistent in every temporary slice and if
the oscillator fluctuations of excitations are permissible 
(just the case of interacting potential above). We imply no fluctuations 
of instanton solution itself (their life-time is of instanton size order).
Let's emphasize that it is worth speaking about the coherent behaviour of
vacuum wave function when the probabilities of  
appearance and disappearance of instantons with different rate of 
size deformation are concordant. Besides, the ensemble projection upon the
three-dimensional space is described by the Lagrangian density 
for every group of the PP's of calibrated sizes (see Eq.(\ref{b1}) 
for partitioning at the initial time)
\begin{equation}
\label{11}
{\cal L}=
-\frac12~\kappa(\rho)~\dot\rho^2 + 
U_{eff}(\rho)~.
\end{equation}

The characteristic scale of excitations is determined at the point of 
potential 
minimum $\frac {d U_{eff}(\rho)}{d\rho}=0$  and is derived from
the equation $\rho_c^{2}=\frac{b-5}{2\nu}~\overline {\rho^2 }.$ Then 
$\frac {d^2 U_{eff}(\rho)}{d\rho^2}\left.\right |_{\rho_c}=
\frac{4\nu}{\overline {\rho^2}}$. Fixing the kinetic energy term (\ref{a9}) 
on  the characteristic scale $\bar\rho$ (similar to the calculation of
potential) one has for the frequency of the proper system fluctuations 
\begin{equation}
\label{mass}
m^2=\frac{4\nu}{\kappa~\overline{\rho^2}}=
\frac{g^2}{8\pi^2}\frac{\nu}{\overline {\rho^2}}=\frac{\nu}
{\beta(\bar\rho)~\overline {\rho^2}}~.
\end{equation}
We have only analyzed the deformations in the temporary direction.
Those in the spatial directions could be estimated by drawing the same
arguments. Thus, the expression for the $\kappa$-term keeps 
the form above obtained with the only change of rates for the
appropriate gradients of function $\rho (x)$ done, i.e. the substitution 
$\dot \rho(t) \to \frac {\partial\rho (x)}{\partial x}$ should be 
performed for such a 'crumpled' instanton. 
Then the frequency of proper fluctuations might be interpreted 
as the mass term and excitations occur having a phonon-like nature
\begin{equation}
\label{12}
{\cal L}=
\frac12~\kappa(\rho)~[-\dot\rho^2+\nabla \rho\nabla \rho]+
U_{eff}(\rho)~,
\end{equation}
(cross-terms $\sim \dot\rho~\rho^{'}$ equal to zero identically).
{\footnote{It is interesting to remark the centre of instanton solution may
not be shifted unlike the dilatational mode, since the relevant
deformations make $\kappa$ singular.}}
The parameters $\bar\rho$ and $\beta(\bar\rho)$ determined 
(by maximizing the partition function of $\bar I I$ liquid 
with respect to N \cite{2})
in a self-consistent way take the following values 
$\bar\rho\Lambda\approx 0.32,~~\beta (\bar\rho)\approx 18.1,
~n~\Lambda^4\approx 0.73,
\footnote{Taken the constant $C_{N_C}$ from Ref. \cite{2} (Nucl. Phys.) as 
$C_{N_c}\approx\frac{4.66~\exp(-1.68 N_c)}{\pi^2(N_c-1)!(N_c-2)!}$
we obtain the following values 
 $\bar\rho~\Lambda\approx 0.37,~~
\beta(\bar\rho)\approx 17.5,~n~\Lambda^4\approx 0.44,$
and for mass gap we have $m\approx 1.21~\Lambda.$}~~(N_c=3$)
therefore, for mass term we have $m\approx 1.37 \Lambda$. 

It is a remarkable fact that instanton liquid model gives 
the effective Lagrangian, incorporating low energy excitations which are 
naturally interpreted as particles. Their quantum numbers are the 
'glueball' ones and the mass scale developed is of light hadron 
mass order. Let's emphasize once more that adiabatic assumption leads to 
fully consistent picture of $\bar I I$ liquid. The model itself regulates
the most suitable regime of dilatation deformations resulting in
the mass gap generation (\ref{mass}). Apparently, an intriguing guess
is to associate the lightest hadrons with these phonon excitations  
discovered, since the preliminary evaluations of their mass spectrum
with the quark condensate included look quite encouraging. 
The concept of confining potential origin for light quarks
in this approach seems simply irrelevant because of stable phonon 
nature.

The Financial support of RFFI: Grants 96-02-16303, 
96-02-00088 G, 
97-02-17491 and INTAS Grant 93-0283, 96-0678 is greatly acknowledged.

\end{document}